\documentclass[a4paper,12pt]{JHEP3}

\usepackage[dvips]{graphicx}
\usepackage{dcolumn}
\usepackage{amsfonts}
\usepackage{latexsym}

\input{colordvi.tex}

\preprint{}

\title{Ghost Dark Matter}

\author{Tomonori Furukawa$^1$, Shuichiro Yokoyama$^1$, Kiyotomo
Ichiki$^1$\\
$^1$Department of Physics and Astrophysics, Nagoya University, Nagoya
464-8602, Japan
}
\author{Naoshi Sugiyama$^{1,2}$, and Shinji Mukohyama$^2$\\
$^2$Institute for the Physics and Mathematics of the Universe (IPMU),
The University of Tokyo, Chiba 277-8582, Japan
}

\date{\today}

\abstract{
We revisit ghost dark matter, the possibility that ghost condensation
may serve as an alternative to dark matter. In particular, we
investigate the Friedmann-Robertson-Walker (FRW) background evolution
and the large-scale structure (LSS) in the $\Lambda$GDM universe,
i.e. a late-time universe dominated by a cosmological constant and
ghost dark matter. The FRW background of the $\Lambda$GDM universe is
indistinguishable from that of the standard $\Lambda$CDM universe if
$M\gtrsim 1~{\rm eV}$, where $M$ is the scale of spontaneous
Lorentz breaking. From the LSS we find a stronger bound:
$M\gtrsim 10~{\rm eV}$. For smaller $M$, ghost dark matter would have
non-negligible sound speed after the matter-radiation equality, and thus
the matter power spectrum would significantly differ from
observation. These bounds are compatible with the phenomenological
upper bound $M\lesssim 100~{\rm GeV}$ known in the literature.
}


\begin{document}                              

\section{Introduction}

Current data of the cosmological observations (e.g., Cosmic Microwave
Background (CMB), Large Scale Structure (LSS) and SuperNovae (SNe)) show
that our Universe today mostly consists of dark matter responsible for
the structure formation and dark energy causing late time accelerated
expansion of the
Universe~\cite{Riess:1998cb,Riess:1998dv,Perlmutter:1998np,Kowalski:2008ez,Hicken:2009dk}. From
observational point of view, the paradigm of dark energy and dark matter
is very successful to fit the data.

However, from theoretical viewpoint, we do not know what they really
are, despite the fact that there are many theoretical models (for
review, see e.g. \cite{Copeland:2006wr} for dark energy and \cite{Bertone:2004pz} for
dark matter). This situation has been a strong motivation for
modification of gravity as an alternative to dark energy and dark
matter: just changing behavior of gravity at long distance/time scales
might be able to explain the observational data without introducing dark
energy and dark matter. Indeed, many theories of modification of gravity
have been proposed, such as massive gravity~\cite{Fierz:1939ix}, DGP
model~\cite{Dvali:2000hr,Deffayet:2000uy}, ghost
condensate~\cite{ArkaniHamed:2003uy,ArkaniHamed:2005gu} and so on.  It
is important to investigate cosmological implications of those modified
gravity theories toward the future observations.

In this paper, we focus on the ghost condensate scenario and investigate
the possibility that it might serve as an alternative to dark
matter. This possibility, dubbed {\it ghost dark matter}, was already
pointed out in \cite{ArkaniHamed:2003uy} but has not been investigated
in detail. Based on the Friedmann-Robertson-Walker (FRW) background
evolution and the large-scale structure of the universe, in the present
paper we shall find a lower bound on the scale of spontaneous Lorentz
breaking, $M\gtrsim 10~{\rm eV}$, under the assumption that ghost dark
matter is responsible for all dark matter in the universe. Most
importantly, this bound is compatible with the phenomenological upper
bound $M\lesssim 100~{\rm GeV}$ found in \cite{ArkaniHamed:2005gu}.

The rest of this paper is organized as follows.  In the next section, we
briefly review the ghost condensate scenario, including the basic idea,
the low-energy effective theory and the phenomenological upper bound on
the scale of spontaneous Lorentz breaking. In Sec.~\ref{sec:background},
we introduce a simplified description of ghost dark matter and
investigate the FRW background evolution. We also clarify the regime of
validity of the simplified description. In Sec.~\ref{sec:LSS}, we
consider effects of ghost dark matter on the large-scale structure of
the universe. We discuss how density perturbations evolve in the
universe dominated by the cosmological constant and ghost dark matter,
i.e. the $\Lambda$GDM universe, and then give a constraint on the model
from the shape of the matter power spectrum. The final section is
devoted to conclusion of this paper and discussions.

\section{Review of ghost condensation}
\label{sec:review}

\subsection{Basic idea}

In particle physics it is the so-called Higgs mechanism that modifies
force law in the infrared (IR) and that makes it possible to describe
the weak interaction in a theoretically controllable way. A
non-vanishing vacuum expectation value (vev) of a scalar field
spontaneously breaks a part of the gauge symmetry and modifies the IR
behavior of the corresponding gauge force from Gauss law to Yukawa law.

Ghost condensation applies the idea of Higgs mechanism to general
relativity to modify the IR behavior of
gravity~\cite{ArkaniHamed:2003uy,ArkaniHamed:2005gu}. In order to
spontaneously break a part of the symmetry of general relativity,
i.e. spacetime diffeomorphism invariance, we consider
{\it a non-vanishing vev of derivative of a scalar field}. In addition,
we demand that {\it the vev is timelike} so that only the time
reparametrization symmetry is spontaneously broken and the
$3$-dimensional spatial diffeomorphism invariance remains unbroken. Note
that this symmetry breaking pattern is totally consistent with
observational fact that our universe averaged over large scales is
isotropic. However, unlike usual situations in cosmology with a scalar
field, we require that {\it the vev of derivative should remain non-zero
and finite as the universe expands towards maximally symmetric
spacetimes}, i.e. Minkowski or de Sitter spacetimes~\footnote{We do not
consider anti-de Sitter spacetimes as backgrounds since we are
interested in cosmology. If we nonetheless considered ghost condensation
in an anti-de Sitter background then the vev of derivative would be
spacelike and excitations around the condensate would not have a healthy
kinetic term.}. This is because we would like the IR modification of
gravity to persist in Minkowski or de Sitter backgrounds.

Therefore the equation of motion of the scalar field and the Einstein
equation must allow a solution with
$X\equiv -\partial^{\mu}\phi\partial_{\mu}\phi$ constant and
positive~\footnote{In this paper we adopt the mostly positive sign for
the spacetime metric.} in Minkowski or de Sitter spacetime. This
requirement forbids inclusion of any non-trivial potentials for the
scalar field since a potential for an eternally running scalar field
would lead to time-dependence in the stress-energy tensor. In other
words, we must invoke {\it the shift symmetry} for the scalar field
action, i.e. invariance of the action under constant shift of the scalar
field: $\phi\to\phi+c$, where $c$ is an arbitrary constant. With the
shift symmetry, the action for the scalar field minimally coupled to
gravity should be of the form
%
\begin{equation}
 I_{\phi} = \int d^4x\sqrt{-g}L_{\phi}, \quad
 L_{\phi}=L_{\phi}(X,\Box\phi,Y,Z,\cdots),
 \label{eqn:general-action}
\end{equation}
where $Y\equiv\nabla^{\mu}\nabla^{\nu}\phi\nabla_{\mu}\nabla_{\nu}\phi$,
$Z\equiv\nabla^{\mu}\nabla^{\nu}\phi\partial_{\mu}\phi\partial_{\nu}\phi$,
and so on. In addition to the shift symmetry, we assume
{\it the $Z_2$ symmetry} for the scalar field action, i.e. invariance of
the action under reflection $\phi\to -\phi$. The $Z_2$ symmetry is to
ensure that the effective theory for excitations of ghost condensate is
invariant under simultaneous reflection of the time $t$ and the scalar
field perturbation $\delta\phi$:
$t\to -t$, $\delta\phi\to -\delta\phi$.

The simple Lagrangian
%
\begin{equation}
 L_{\phi} = P(X) \label{eqn:actionP}
\end{equation}
is of the form (\ref{eqn:general-action}). The equation of motion
for a homogeneous $\phi(t)$ in the flat Friedmann-Robertson-Walker (FRW)
background
%
\begin{equation}
 ds^2 = -dt^2 + a(t)^2(dx^2+dy^2+dz^2)
\end{equation}
is $\partial_t(a^3P_X\partial_t\phi)=0$, where $P_X\equiv dP/dX$, and
leads to
%
\begin{equation}
 P_X X^{1/2} \propto a^{-3} \to 0 \quad (a\to \infty),
\end{equation}
i.e. either $P_X\to 0$ or $X\to 0$. Note that $X$ is either positive or
zero for a homogeneous $\phi$. Thus, if $P(X)$ has an extremum at
$X=M^4>0$, i.e. $P_X(M^4)=0$, then $X=M^4$ is a dynamical attractor of
the system and ghost condensate may be realized automatically.

However, we shall see below that a more general action depending on
e.g. $\Box\phi$ (see (\ref{eqn:general-action})) is needed to describe
excitations around the ghost condensate properly.

\subsection{Decoupling limit in Minkowski background}
\label{subsec:GCMin}

Let us now consider excitations around the extremum $X=M^4>0$ of
$P(X)$. For simplicity we consider Minkowski background. We shall
expand the action (\ref{eqn:actionP}) around $\phi=c t$ and take the
limit $c^2\to M^4$. Note that in Minkowski background the equation of
motion for $\phi$ is satisfied for any $c$ and that this treatment is
consistent as far as the backreaction to the geometry is negligible.

For
%
\begin{equation}
 \phi = c t + \pi(t,\vec{x}),
\end{equation}
by expanding $P(X)$ with respect to $\pi$ we obtain the quadratic
Lagrangian for $\pi$ as
%
\begin{equation}
 L_{\pi}^{(0)} = \left[2c^2P_{XX}(c^2)+P_X(c^2)\right](\partial_t\pi)^2
 - P_X(c^2)(\vec{\nabla}\pi)^2,
 \label{eqn:Lpi0}
\end{equation}
where $P_{XX}\equiv d^2P/dX^2$. Note that we did not take into account
the backreaction of the scalar field to the background geometry nor
include metric perturbations. These treatments are justified in the
limit $E/M_{\rm Pl}\to 0$, where $E$ represents energy scales of
interest and $M_{\rm Pl}$ is the Planck scale. Since we shall later take
the limit $c^2\to M^4$ and we shall see that $M$ sets the cutoff scale
of the effective field theory, $E$ can be replaced by $M$ in the regime
of validity of the effective field theory. Thus, the decoupling limit is
characterized by $M/M_{\rm Pl}\to 0$. In this limit and for energies and
momenta sufficiently lower than $M$, the action (\ref{eqn:Lpi0}) is
valid and the small fluctuation $\pi$ is stable if
%
\begin{equation}
 2c^2P_{XX}(c^2)+P_X(c^2) > 0, \quad
 P_X(c^2) > 0.  \label{eqn:stabilitycondition}
\end{equation}
By taking the limit $c^2\to M^4$, we obtain
%
\begin{equation}
 L_{\pi}^{(0)} = 2M^4P_{XX}(M^4)(\partial_t\pi)^2.
 \label{eqn:timekineticterm}
\end{equation}
Thus, the excitation $\pi$ around the attractor $X=M^4$ has a healthy
time kinetic term if $P_{XX}(M^4)>0$, i.e. if $X=M^4$ is a local
minimum of the function $P(X)$. However, the spatial gradient term
vanishes.

This means that the action (\ref{eqn:actionP}) is too simple to describe
excitations around the ghost condensate background. Indeed, while we
have assumed the shift symmetry to prevent a non-trivial potential from
being generated, there is no way to prevent $\Box\phi$, $Y$, $Z$,
etc. from appearing in the action. We therefore have to go back to the
general action (\ref{eqn:general-action}) and seek leading gradient
terms.

It turns out that the leading gradient term is of the form
%
\begin{equation}
 \Delta L_{\pi} = -\frac{\alpha}{2M^2}(\vec{\nabla}^2\pi)^2
 \label{eqn:gradientterm}
\end{equation}
where $\alpha$ is a constant of order unity. (The reason why this is
indeed the leading term will be made clear in the next subsection.)
Combining this with the leading time kinetic term
(\ref{eqn:timekineticterm}), we obtain the quadratic action for $\pi$
%
\begin{equation}
 I_{\pi} = \int dt d^3\vec{x}L_{\pi}, \quad
 L_{\pi} = M^4\left\{\frac{1}{2}(\partial_t\pi)^2
 -\frac{\alpha}{2M^2}(\vec{\nabla}^2\pi)^2\right\},
 \label{eqn:quadraticaction}
\end{equation}
where we have normalized $\pi$ and $\alpha$ as
$2\sqrt{P_{XX}(M^4)}\pi\to\pi$ and
$\alpha\to 4M^4P_{XX}(M^4)\alpha$. Thus, the dispersion relation for
$\pi$ in the decoupling limit $M/M_{\rm Pl}\to 0$ is
%
\begin{equation}
 \omega^2 = \frac{\alpha}{M^2}\vec{k}^4.
 \label{eqn:dispersionrelationdecoupling}
\end{equation}

\subsection{Scaling dimension and suppression of extra terms}
\label{subsec:scalingdim}

The general action (\ref{eqn:general-action}) should in principle
include any terms consistent with the shift symmetry and the $Z_2$
symmetry. In terms of $\pi$, they are invariance under the constant
shift of $\pi$ ($\pi\to\pi+c$, where $c$ is an arbitrary constant) and
invariance under the simultaneous reflection of the time $t$ and $\pi$
($t\to -t$, $\pi\to -\pi$), respectively. Thus, there should be infinite
number of terms added to the quadratic action
(\ref{eqn:quadraticaction}) for $\pi$. Nonetheless, one can show that
the quadratic action (\ref{eqn:quadraticaction}) is a good description
of low energy behavior of the system.

In order to show that those extra terms are irrelevant at low energies,
let us first identify the scaling dimensions as the energy $E$ is scaled
by $E\to sE$ (or the time interval $dt$ is scaled by $dt\to s^{-1}dt$),
where $s$ is some constant. By requiring that the quadratic action
(\ref{eqn:quadraticaction}) be invariant under scaling, we can fix the
scaling dimensions as
%
\begin{eqnarray}
 E & \to & sE, \nonumber\\
 dt & \to & s^{-1}dt, \nonumber\\
 d\vec{x} & \to & s^{-1/2}d\vec{x}, \nonumber\\
 \pi & \to & s^{1/4}\pi.
\end{eqnarray}
Note that the scaling dimensions of $dt$ and $d\vec{x}$ are consistent
with the dispersion relation
(\ref{eqn:dispersionrelationdecoupling}). With this scaling, one can
check that the leading interaction
%
\begin{equation}
 M^4 \int dt d^3\vec{x} (\vec{\nabla}\pi)^2\partial_t\pi
\end{equation}
scales as $s^{1/4}$. Thus, this term is irrelevant and becomes less and
less important at energies and momenta sufficiently lower than $M$. All
other terms are even more irrelevant.

There is one relevant operator, namely $(\vec{\nabla}\pi)^2$. However,
as we have already seen, the coefficient of this operator is
proportional to $P_X$ and goes to zero as the universe expands.

Therefore, if energies, momenta and the field amplitude are sufficiently
lower than $M$ then low energy/momentum/amplitude expansion around the
quadratic action (\ref{eqn:quadraticaction}) is under control. This in
particular implies that apparent extra modes due to higher time
derivative terms have frequencies of order $M$ or higher. Hence there is
no ghost in the regime of validity of the effective field theory if we
set the cutoff scale slightly below $M$.

\subsection{Jeans-like instability and IR modification of linear gravity}

So far, we have considered the decoupling limit $M/M_{\rm Pl}\to 0$ of
the theory. For small but finite $M/M_{\rm Pl}$, the dispersion relation
(\ref{eqn:dispersionrelationdecoupling}) gets corrected due to mixing
with gravity and becomes~\cite{ArkaniHamed:2003uy}
%
\begin{equation}
 \omega^2 = \frac{\alpha}{M^2}\vec{k}^4
 - \frac{\alpha M^2}{2M_{\rm Pl}^2}\vec{k}^2.
 \label{eqn:dispersionrelation}
\end{equation}
This dispersion relation exhibits Jeans-like instability for modes with
length scales longer than $L_c$ and the corresponding time scale is
$T_c$, where
%
\begin{equation}
 L_c \sim \frac{M_{\rm Pl}}{M^2}, \quad
 T_c \sim \frac{M_{\rm Pl}^2}{\sqrt{\alpha}M^3}.
\end{equation}
Note that this is an IR instability and has nothing to do with
ghost. The Jeans-like instability is the origin of the IR modification
of gravity in ghost condensate background~\cite{ArkaniHamed:2003uy}. The
time scale and the length scale of the modification are $T_c$ and $L_c$,
respectively, and are much longer than the naive scale $1/M$.

In Sec.~\ref{sec:LSS} we shall investigate Jeans instability of ghost
dark matter, a component which arises from excitation around ghost
condensate and which behaves like dark matter. One should note that the
Jeans-like instability in the exact ghost condensate background
considered in this subsection is both conceptually and qualitatively
different from the Jeans instability of ghost dark matter in
Sec.~\ref{sec:LSS}.

In the decoupling limit $M/M_{\rm Pl}\to 0$, the timescale $T_c$
diverges (in the unit of $1/M$) and thus the Jeans-like instability
disappears.

If we required that the Jeans timescale $T_c$ be longer than the age of
the universe then we would end up with the (would-be) upper bound
$M\lesssim 10~{\rm MeV}$~\cite{ArkaniHamed:2003uy}. However, we shall see
below that nonlinear dynamics becomes important much earlier than $T_c$,
that this (would-be) bound is not necessary and that the current upper
bound on $M$ is as weak as
$M\lesssim 100~{\rm GeV}$~\cite{ArkaniHamed:2005gu}.

\subsection{Nonlinear dynamics and upper bound on $M$}
\label{subsec:nonlinear}

A sightly nonlinear extension of the quadratic action
(\ref{eqn:quadraticaction}) coupled to linearized gravity is
%
\begin{equation}
 I_{\pi} = \int dtd^3\vec{x}L_{\pi}, \quad
 L_{\pi} = M^4\left\{\frac{1}{2}
 \left[\partial_t\pi-(\vec{\nabla}\pi)^2-\Phi\right]^2
 - \frac{\alpha}{2M^2}(\vec{\nabla}^2\pi)^2\right\},
 \label{eqn:nonlinearaction}
\end{equation}
where $\Phi=-\delta g_{00}/2$ is the Newtonian
potential~\cite{ArkaniHamed:2005gu} in the longitudinal gauge. This is
obtained by not dropping the leading nonlinear term and the metric
perturbation in going from (\ref{eqn:actionP}) to
(\ref{eqn:timekineticterm}) before adding (\ref{eqn:gradientterm}). Note
that $(\partial_t\pi)^2$ has been dropped from the expression in the
square bracket since it has a scaling dimension higher than
$(\vec{\nabla}\pi)^2$, following the discussions in
subsection~\ref{subsec:scalingdim}.

From the quadratic part of the action (\ref{eqn:nonlinearaction}),
i.e. (\ref{eqn:quadraticaction}), it is easy to see that the timescale
of linear dynamics $T_{\rm Lin}$ is determined by
%
\begin{equation}
 \frac{\pi^2}{T_{\rm Lin}^2} \sim \ \frac{\pi^2}{M^2L^4},
\end{equation}
where we have assumed that the length scale of interest $L$ is shorter
than the Jeans scale $L_J$ (so that the second term in the right hand
side of (\ref{eqn:dispersionrelation}) is negligible) and we have set
$\alpha=O(1)$. Thus we obtain
%
\begin{equation}
 T_{\rm Lin} \sim M L^2,
\end{equation}
which is consistent with the dispersion relation
(\ref{eqn:dispersionrelationdecoupling}). On the other hand, from the
nonlinear action (\ref{eqn:nonlinearaction}), the timescale of
nonlinear dynamics $T_{\rm NL}$ is determined by
%
\begin{equation}
 \frac{\pi}{T_{\rm NL}} \sim \ \frac{\pi^2}{L^2} \sim \Phi,
\end{equation}
and we obtain
%
\begin{equation}
 T_{\rm NL} \sim \frac{L}{\sqrt{|\Phi|}}
 \sim \sqrt{\frac{M_{\rm Pl}^2L^3}{M_{\rm src}}},
\end{equation}
where $M_{\rm src}$ is the mass of the gravitational source. Note that
this timescale is nothing but the Kepler time. Therefore, nonlinear
dynamics dominates before linear dynamics if
$T_{\rm NL}\lesssim T_{\rm Lin}$, i.e.
%
\begin{equation}
 \frac{M}{M_{\rm Pl}} \gtrsim \sqrt{\frac{1}{M_{\rm src}L}}.
\end{equation}
This condition is satisfied in virtually all interesting
situations. For example, for the earth's surface gravity, this condition
is as weak as
%
\begin{equation}
 M \gtrsim 10^{-9}~{\rm eV}.
\end{equation}

From the dispersion relation (\ref{eqn:dispersionrelation}) it is easy
to see that modes with $L\gtrsim L_J$ grow due to Jeans-like instability
in timescale
%
\begin{equation}
 \tau \sim \frac{M_{\rm Pl}}{M}L.
\end{equation}
The nonlinear term $(\vec{\nabla}\pi)^2$ in the squared bracket in the
action (\ref{eqn:nonlinearaction}) becomes important when it is
comparable to $\partial_t\pi$ or larger, i.e. when
%
\begin{equation}
 |\pi| \gtrsim \pi_c \equiv \frac{L^2}{\tau}.
\end{equation}
This can be rewritten as a condition on the energy density
$\rho_{\pi}\sim M^4\partial_t\pi$:
%
\begin{equation}
 \rho_{\pi} \gtrsim \rho_c \equiv \frac{M^4\pi_c}{\tau}
 \sim \frac{M^6}{M_{\rm Pl}^2}.
\end{equation}
Hereafter, we assume that nonlinear dynamics cutoff the Jeans-like
instability at $|\rho_{\pi}|\sim\rho_c$. Positive and negative regions
can grow to $|\rho_{\pi}|\sim\rho_c$ within the age of the universe
$H_0^{-1}$ if $L\gtrsim L_c$ and $\tau\lesssim H_0^{-1}$, i.e. if
%
\begin{equation}
 L_c \lesssim L \lesssim L_{\rm max} \equiv \frac{M}{M_{\rm Pl}H_0}
 \sim 10 R_{\odot}\times\frac{M}{100{\rm GeV}},
\end{equation}
where $R_{\odot}$ is the solar radius. Therefore, if $M$ is lower than
$100~{\rm GeV}$ (we shall indeed see below that $M$ must be lower than
$100~{\rm GeV}$) then the typical size of positive and negative regions
is smaller than $10R_{\odot}$. This means that $\rho_{\pi}$ averaged
over large scales such as galactic scales is almost zero and does not
gravitate significantly.

Based on these properties of nonlinear dynamics, several
phenomenological upper bounds on $M$ were derived in
\cite{ArkaniHamed:2005gu}. The strongest among them is the
'twinkling-from-lensing' bound, which we shall briefly describe
here.

Suppose that the universe is filled with regions with
$\rho_{\pi}\sim\pm\rho_c$ of the size
$L_c\lesssim L\lesssim L_{\rm max}$ moving relative to the cosmic
microwave background (CMB). Each region contributes to weak
gravitational lensing with the deflection angle
%
\begin{equation}
 \Delta\theta_{\rm each} \sim \frac{r_g}{b}
 \sim \frac{M^6b^2}{M_{\rm Pl}^4}
 \sim \frac{M^6L^2}{M_{\rm Pl}^4},
\end{equation}
where $b$ ($\lesssim L$) is the impact parameter and
$r_g\sim \rho_c b^3/M_{\rm Pl}^2$ is the gravitational radius of the
mass contained within the impact parameter. In the final expression, we
have maximized $\Delta\theta_{\rm each}$ with respect to $b$. Since a
light-ray from the distance $d$ experiences $N$ ($\sim d/L$) lens
events, the total deflection angle is
%
\begin{equation}
 \Delta\theta_{\rm tot} \sim \Delta\theta_{\rm each}\sqrt{N}
 \sim \frac{M^6d^{1/2}L^{3/2}}{M_{\rm Pl}^4}
 \sim \frac{M^6d^{1/2}L_{\rm max}^{3/2}}{M_{\rm Pl}^4},
\end{equation}
where we have maximized $\Delta\theta_{\rm tot}$ with respect to $L$ in
the final expression. We can apply this result to the CMB by setting
$d\sim H_0^{-1}$. Requiring that $\Delta\theta_{\rm tot}$ be smaller
than the angular resolution of CMB experiments $\sim 10^{-3}$, we obtain
the upper bound
%
\begin{equation}
 M\lesssim 100~{\rm GeV}.
 \label{eqn:upperbound}
\end{equation}
The twinkling timescale for the CMB is
%
\begin{equation}
 T_{\rm twinkle} \sim \frac{L_{\max}}{v}
 \sim \frac{M}{100{\rm GeV}}\cdot
 \frac{300{\rm km/s}}{v}\times 0.1~ {\rm day},
\end{equation}
where $v$ is the typical velocity of positive and negative regions
relative to the CMB rest frame. Thus, if $M$ is close to $100~{\rm GeV}$
then the twinkling effect may be detected in future CMB experiments.

\section{Ghost dark matter and background evolution}
\label{sec:background}

\subsection{Simple description of ghost dark matter}
\label{subsec:simple-gdm}

Now let us reconsider the simple Lagrangian (\ref{eqn:actionP}) which
depends on $X$ only. As we have already seen in
subsection~\ref{subsec:GCMin}, we need to add the extra term
(\ref{eqn:gradientterm}) to this Lagrangian in order to describe
perturbations around the exact ghost condensate background, i.e. a local
minimum of the function $P(X)$. In this subsection we shall instead
consider a background with non-vanishing $P_X$ and see that, under a
certain condition, the Lagrangian $P(X)$ without the additional term
(\ref{eqn:gradientterm}) can properly describe the background and
perturbations around it. In this situation, as we shall see below,
deviation from the exact ghost condensate behaves like dark matter.

The stress-energy tensor corresponding to the Lagrangian $P(X)$ is
%
\begin{equation}
 T_{\mu\nu} =
 2P_X\partial_{\mu}\phi\partial_{\nu}\phi + Pg_{\mu\nu}
 = (\rho+P)u_{\mu}u_{\nu}+Pg_{\mu\nu},
\end{equation}
where
%
\begin{equation}
 \rho = 2XP_X-P, \quad u_{\mu} = -\frac{\partial_{\mu}\phi}{\sqrt{X}}.
\end{equation}
The sound speed squared for perturbation is~\cite{Garriga:1999vw}
%
\begin{equation}
 c_s^2 = \frac{dP/dX}{d\rho/dX} = \frac{P_X}{2XP_{XX}+P_X}.
 \label{eqn:soundspeedsquared}
\end{equation}
This agrees with minus the ratio of the coefficient of the gradient term
to the coefficient of the time kinetic term in the quadratic Lagrangian
(\ref{eqn:Lpi0}).

As we have already seen in subsection~\ref{subsec:GCMin}, in order
to describe perturbations around the exact ghost condensate background,
the simple Lagrangian $P(X)$ is not sufficient but we need to add the
term (\ref{eqn:gradientterm}) to it. The reason is that the coefficient
of the would-be leading gradient term $(\vec{\nabla}\pi)^2$ vanishes in
the exact ghost condensate background. However, if the background is
not exact ghost condensate but has $P_X\ne 0$ then the coefficient
of $(\vec{\nabla}\pi)^2$ does not vanish as shown in
(\ref{eqn:soundspeedsquared}). Therefore, if $c_s^2$ is large enough
then we do not have to add the extra term (\ref{eqn:gradientterm}) to
the simple Lagrangian $P(X)$. To be more precise, the extra term
(\ref{eqn:gradientterm}) is not needed if
%
\begin{equation}
 c_s^2 \gg \frac{1}{M^2L^2},
 \label{eqn:largeenoughcs}
\end{equation}
where $L$ is the length scale of interest and we have supposed that
$\alpha=O(1)$.

We have also seen that a local minimum of $P(X)$ is a dynamical
attractor in the expanding universe. Thus, it is rather natural to
Taylor expand $P(X)$ around a local minimum $X=M^4$ as
%
\begin{equation}
 P \simeq P(M^4) + \frac{1}{2}P_{XX}(M^4)(X-M^4)^2.
\end{equation}
This expansion of $P(X)$ is valid if and only if $|X-M^4|\ll M^4$. Note
that we anyway have to restrict our consideration to this regime;
otherwise, higher dimensional operators, which in general depend on not
only $X$ but also $\Box\phi$, $Y$, $Z$, etc., would be unsuppressed and
the system would exit the regime of validity of the low energy effective
theory. (See discussions in subsection \ref{subsec:scalingdim}.)
Correspondingly, we have the following expansions.
%
\begin{eqnarray}
 \rho & \simeq & -P(M^4) +2M^4P_{XX}(M^4)(X-M^4), \label{eqn:rhogdm}
 \\
 c_s^2 & \simeq & \frac{X-M^4}{2M^4}. \label{eqn:soundspeed}
\end{eqnarray}
These expressions are rewritten as
%
\begin{equation}
 \rho \simeq \rho_{\rm gde} + \rho_{\rm gdm}, \quad
 P \simeq P_{\rm gde} + P_{\rm gdm},
 \quad c_s^2 \simeq \frac{\rho_{\rm gdm}}{\bar{M}^4}
 \label{eqn:GDE-GDM1}
\end{equation}
where
%
\begin{equation}
 P_{\rm gde} = -\rho_{\rm gde} = \mbox{const.}, \quad
 P_{\rm gdm}=\frac{\rho_{\rm gdm}^2}{2\bar{M}^4}.
 \label{eqn:GDE-GDM2}
\end{equation}
Here, $\bar{M}^4\equiv 4M^8P_{XX}(M^4)\sim M^4$. In the regime
$|X-M^4|\ll M^4$, it is intriguing to note that ($\rho_{\rm gde}$,
$P_{\rm gde}$) and ($\rho_{\rm gdm}$, $P_{\rm gdm}$) behave like dark
energy and dark matter, respectively. In particular, we shall call the
latter {\it ghost dark matter}~\cite{ArkaniHamed:2005gu}.

In terms of ghost dark matter component, the necessary condition
(\ref{eqn:largeenoughcs}) for the validity of the simple Lagrangian
$P(X)$ is written as
%
\begin{equation}
 M \ll L\rho_{\rm gdm}^{1/2} \simeq 10^{15}{\rm GeV}
 \times \left(\frac{L}{1{\rm Mpc}}\right)\cdot
 \left(\frac{\rho_{\rm gdm}}{0.3\times\rho_0}\right)^{1/2},
 \label{eqn:validityGDM1}
\end{equation}
where $\rho_0\equiv 3M_{\rm Pl}^2H_{0}^2$ is the critical
density. For example, if $M\ll 10^{15}~{\rm GeV}$ and if we
suppose that the ghost dark matter is responsible for all dark matter in
the universe ($\rho_{\rm gdm}\simeq 0.3\times\rho_0$) then
(\ref{eqn:validityGDM1}) is satisfied for length scales longer than
$\sim 1~{\rm Mpc}$. Note, however, that validity of low energy effective
theory requires that all associated energies, momenta and amplitudes be
sufficiently lower than unity in the unit of $M$. Thus,
(\ref{eqn:GDE-GDM1}) and (\ref{eqn:GDE-GDM2}) are valid only if
%
\begin{equation}
 \rho_{\rm gdm} \ll M^4, \quad H\ll M.   \label{eqn:validityGDM2}
\end{equation}
On the other hand, we do not have to require that the radiation
temperature be lower than $M$ since interactions between ghost
condensate and radiation are highly suppressed (typically by the Planck
scale).

\subsection{Ghost dark matter production from ghost inflation}
\label{subsec:gi2gdm}

In the previous subsection we have seen that the background evolution
and the behavior of perturbations of ghost dark matter can be described
by (\ref{eqn:GDE-GDM1}) and (\ref{eqn:GDE-GDM2}) under the conditions
(\ref{eqn:validityGDM1}) and (\ref{eqn:validityGDM2}), where $L$ is the
length scale for perturbations of interest and $\bar{M}\sim M$. If
(\ref{eqn:validityGDM1}) is not satisfied then the use of the simple
Lagrangian of the form $P(X)$ is not justified and we need to take into
account effects of the extra term (\ref{eqn:gradientterm}). This is not
a big problem but would make analysis slightly
complicated~\cite{Mukohyama:2006be}. Fortunately, for the purpose of the
present paper, i.e. for understanding of the evolution of the FRW
background and the large-scale structure of the universe, we are
interested in $L$ of order $1~{\rm Mpc}$ or longer. In this case, the
condition (\ref{eqn:validityGDM1}) always holds if the phenomenological
upper bound $M\lesssim 100~{\rm GeV}$ (see
subsection~\ref{subsec:nonlinear}) is satisfied and if we suppose that a
non-trivial fraction of dark matter of the universe is ghost dark
matter. On the other hand, if (\ref{eqn:validityGDM2}) is not met then
the system exits the regime of validity of the low energy effective
theory and we need a ultraviolet (UV) completion~\footnote{See
\cite{Graesser:2005ar,O'Connell:2006de,Mukohyama:2006mm,Bilic:2008pe}
for some attempts towards possible scenarios of UV completion, and
\cite{Mukohyama:2009rk,Mukohyama:2009um}
for compatibility with the generalized second law of black hole
thermodynamics.} to describe the system.

This naturally leads to the question ``what happens in the early
universe?'' The condition (\ref{eqn:validityGDM2}) does not hold in the
very early epoch of the universe. In this early epoch the sector
including ghost condensation should be governed by a theory more
fundamental than what we have been describing since $M$ is the energy
scale above which a new physics kicks in. While it is important to seek
a UV completion to describe this epoch properly, it is also plausible to
consider cosmological scenarios in which all interesting observables are
predicted within the regime of validity of the low energy effective
theory. As a possible realization of such scenarios let us consider a
generation mechanism of ghost dark matter at the end of ghost
inflation~\cite{ArkaniHamed:2003uz}.

In ghost inflation the scalar field $\phi$ responsible for ghost
condensation plays the role of inflaton as well. For example we can
consider a hybrid inflation-type implementation. In this case we suppose
that the mass squared $m_{\chi}^2$ of a water-fall field $\chi$ depends
on $\phi$ in such a way that $m_{\chi}^2(\phi)$ is positive and constant
for $\phi\ll-\phi_*$, and negative and constant for $\phi\gg\phi_*$,
where $\phi_*$ is a positive constant. This setup is technically natural
since the shift symmetry is broken only in the vicinity of the
transition region $|\phi|\lesssim\phi_*$ and otherwise exact. We suppose
that $\partial_t\phi>0$ so that the sign of $m_{\chi}^2$ changes from
positive to negative.

For $\phi\ll-\phi_*$, $\phi$ enjoys the shift symmetry and $\chi$ has a
positive mass squared. Thus the system has a de Sitter attractor with
$\partial_t\phi=M^2$ and $\chi=0$, where $X=M^4$ is a local minimum of
$P(X)$. We suppose that the system settles in this state well before
$\phi$ crosses the transition region.
Noting that the fluctuation $\delta\phi$ of $\phi$ has the scaling
dimension $1/4$ (see subsection~\ref{subsec:scalingdim}) and the mass
dimension $1$, we can easily estimate the amplitude of quantum
fluctuations of $\phi$ as
%
\begin{equation}
 \delta\phi \sim M\left(\frac{H_{\rm inf}}{M}\right)^{1/4},
\end{equation}
where $H_{\rm inf}$ is the Hubble expansion rate of the de Sitter
attractor. As usual, quantum fluctuations of $\phi$ is eventually
converted to temperature anisotropies as
%
\begin{equation}
 \frac{\delta T}{T} \sim
 \frac{H_{\rm inf}\delta\phi}{\partial_t\phi}.
\end{equation}
Thus, we obtain
%
\begin{equation}
 \frac{\delta T}{T} \sim
 \left(\frac{H_{\rm inf}}{M}\right)^{5/4}.
\end{equation}
By requiring that this is responsible for the observed amplitude of
temperature anisotropies $\delta T/T\sim 10^{-5}$,
$H_{\rm inf}$ is determined as
%
\begin{equation}
 H_{\rm inf} \simeq 10^{-4}\times M.
 \label{eqn:Hinf}
\end{equation}
Under this condition, one can also estimate non-Gaussian features of
CMB anisotropies. The shape of the bispectrum is of the equilateral
type and the nonlinear parameter $f_{NL}$ is of order $\sim 80$ if we
set all dimensionless parameters to unity~\cite{ArkaniHamed:2003uz}. The
essential reason for the relatively large non-Gaussianities is that the
leading nonlinear operator has the scaling dimension $1/4$ and thus is
less suppressed than in usual slow-roll inflation.

The condition (\ref{eqn:Hinf}) shows that ghost inflation is well within
the regime of validity of the low energy effective theory:
$H_{\rm inf}\ll M$. The condition (\ref{eqn:Hinf}) also shows that there
is a lower bound on $M$ in terms of the reheating temperature
$T_{\rm reh}$:
%
\begin{equation}
 M \gtrsim 10^4\times \frac{T_{\rm reh}^2}{M_{\rm Pl}}.
 \label{eqn:bound-inflation}
\end{equation}

Ghost inflation can generate not only temperature anisotropies observed
in the CMB but also ghost dark matter. In the hybrid inflation-type
implementation, the shift symmetry is broken and a potential for $\phi$
should be generated by quantum corrections in and only in the vicinity
of the transition region $|\phi|\lesssim\phi_*$. Correspondingly, the
equation of motion for homogeneous $\phi$ is
%
\begin{equation}
 \frac{1}{a^3}\partial_t\left[a^3P_X\partial_t\phi\right]
 + V'_{\rm gen}(\phi) = 0, \label{eqn:EOMwithVgen}
\end{equation}
where $V_{\rm gen}(\phi)$ is the potential for $\phi$ generated by
quantum corrections. Note that it is not $V_{\rm gen}(\phi)$ but a
potential for $\chi$ that is responsible for most of the potential
energy during the inflationary phase:
%
\begin{equation}
 |\Delta V_{\rm gen}| \ll 3M_{\rm Pl}^2H_{\rm inf}^2,
 \label{eqn:DeltaVgensmall}
\end{equation}
where
$\Delta V_{\rm gen}\equiv
V_{\rm gen}^{\rm before}-V_{\rm gen}^{\rm after}$,
and the superscripts 'before' and 'after' represent values before and
after the transition. Noting that $\partial_t\phi\simeq M^2$ and
ignoring the Hubble friction term, it is easy to integrate
(\ref{eqn:EOMwithVgen}) over the transition region. The result is
%
\begin{equation}
 P_X^{\rm after} \simeq \frac{\Delta V_{\rm gen}}{M^4}.
 \label{eqn:PXafter}
\end{equation}
This gives the amplitude of ghost dark matter right after the transition
as
%
\begin{equation}
 \rho_{\rm gdm}^{\rm after} \sim \Delta V_{\rm gen}.
\end{equation}
Ghost dark matter does not dominate the universe just after the end of
ghost inflation as easily seen from (\ref{eqn:DeltaVgensmall}). However,
since it evolves more slowly than radiation, it can dominate the late
time universe.

In obtaining the estimate (\ref{eqn:PXafter}) we have ignored the
Hubble friction. This is justified if the transition timescale,
$\Delta t\sim\phi_*/\partial_t\phi\simeq\phi_*/M^2$, is short compared
with the Hubble timescale $H_{\rm inf}^{-1}\simeq 10^4/M$, i.e. if
%
\begin{equation}
 \phi_* \ll 10^4\times M.
\end{equation}

\subsection{Phenomenological constraint from background evolution}
\label{subsec:background}

While it is certainly interesting and important to investigate concrete
cosmological scenarios such as one presented in the previous subsection,
it is also important to constrain the ghost dark matter in a model
independent way. This subsection and the next section are devoted to
this subject, assuming that ghost dark matter is responsible for all
dark matter in the universe.

We now consider background evolution from the radiation-matter equality
to the present time to give a lower bound on $M$. In this epoch we know
from observations that the background dark matter component behaves like
pressure-less dust. Thus, if all dark matter in the universe is ghost
dark matter then
$P_{\rm gdm}/\rho_{\rm gdm}\sim \rho_{\rm gdm}/M^4$ must be sufficiently
lower than unity. (This condition incidentally agrees with the first
inequality in (\ref{eqn:validityGDM2}).) In this case $\rho_{\rm gdm}$
behaves as
%
\begin{equation}
 \rho_{\rm gdm} \simeq
 \Omega_{\rm gdm}\rho_0\cdot\left(\frac{a}{a_0}\right)^{-3}.
\end{equation}
We set $\Omega_{\rm gdm}\simeq 0.3$ by the assumption that ghost
dark matter is responsible for all dark matter in the universe. Hence,
by requiring that $\rho_{\rm gdm}\ll M^4$ all the way up the
matter-radiation equality $a\simeq 10^{-4}a_0$, we obtain the lower
bound
%
\begin{equation}
 M \gtrsim 1~{\rm eV}.
\end{equation}


\section{Large-scale structure with ghost dark matter}
\label{sec:LSS}

In this section, we calculate the evolution of density perturbations
in the $\Lambda$GDM universe,  where the sound speed for matter
perturbation, $c_s$, is given by Eq.~(\ref{eqn:soundspeed}) unlike the
standard $\Lambda$CDM universe. Then, we consider the bound on $M$ in
order not to conflict with current observations of large-scale
structure.

\subsection{Jeans wavenumber for ghost dark matter}

In subsection \ref{subsec:background}, we obtained the lower bound on $M$
as $M\gtrsim 1~{\rm eV}$ by demanding that
$P_{\rm gdm}/\rho_{\rm gdm}\ll 1$ all the way up to the matter-radiation
equality. The condition $M\gtrsim 1~{\rm eV}$ also justifies the use of
the simplified description given in subsection \ref{subsec:simple-gdm}
for the matter dominated era since $P_{\rm gdm}/\rho_{\rm gdm}\ll 1$ is
equivalent to $\rho_{\rm gdm}\ll M^4$.

We consider the evolution of the matter perturbations in this
situation. In this subsection, before detailed numerical analysis in the
next subsection, let us foresee that a bound stronger than
$M\gtrsim 1~{\rm eV}$ can be obtained.

For the sound speed given by Eq.~(\ref{eqn:soundspeed}),
we can define the Jeans wavenumber as
\begin{eqnarray}
k_J = \sqrt{{3 \over 2}}{aH \over c_s}~.
\label{eq:Jeans}
\end{eqnarray}
As is well-known, the matter perturbations do not evolve very much
during radiation dominated era and we need not to consider the effects
of the Jeans scales.  However, after the matter-radiation equality we
have to take into account the effects of the Jeans scales, namely, the
perturbations with scales shorter than the Jeans scales can not grow.
From Eq.~(\ref{eq:Jeans}), the Jeans wavenumber evolves as $k_J \propto
a$ during the matter dominated era and hence in the matter dominated era
the comoving Jeans scale ($\sim k_J^{-1}$) becomes the largest at the
matter-radiation equality.  The Jeans wavenumber at the matter-radiation
equality is given by
\begin{eqnarray}
k_{J,{\rm eq}}
\simeq 1 \times
\left( {\Omega_{\rm gdm}h^2 \over 0.11} \right)^{-5/6}
\left( {M \over 10 {\rm eV} } \right)^{4/3}~{\rm Mpc}^{-1}.
\label{eqn:critwavenumber}
\end{eqnarray}
The matter power spectrum is expected to be significantly suppressed for
modes with wavenumbers $k\gtrsim k_{J,{\rm eq}}$.

Almost all current observational data do not indicate any suppression
in the matter power spectrum,
roughly for the wavenumber $k/h \lesssim 1$ Mpc$^{-1}$.
Hence, from the above expression,
we can roughly obtain  a constraint on the model parameter $M$ as
$M \gtrsim 10$ eV.
In the next subsection, we shall confirm this bound by numerically
calculating the matter power spectrum.

\subsection{Numerical calculation}
\subsubsection{Field description and background evolution}
{
In the numerical calculation, it will be practically difficult to use
Eqs.~(\ref{eqn:rhogdm})-(\ref{eqn:GDE-GDM2}) because the sound speed of
GDM increases beyond the speed of light as we go back in time.
(Of course, this is not a physical problem but just a breakdown of the
description.)
Therefore, in the following, we shall solve $X$ assuming
a working Lagrangian to avoid this difficulty and follow the time evolution in a
numerically stable way, instead of solving
Eqs.~(\ref{eqn:rhogdm})-(\ref{eqn:GDE-GDM2}).
Let us consider a model in which $P(X)$ is given by~\footnote{
In Ref. \cite{Giannakis:2005kr}, the authors have also investigated
cosmological implications for the dark matter model whose Lagrangian
is given by Eq. (4.3), called "kinetic UDM" model in that paper.
}
\begin{equation}
P(X)=\frac{1}{8M^4}(X-M^4)^2~.
\end{equation}
In this case $\bar{M}=M$. The time evolution of $X$ is given by
\begin{equation}
\frac{dX}{d\ln a} = \frac{6X(X-M^4)}{3X-M^4}~,\label{eq:4.6}
\end{equation}
and the energy density, pressure, and sound speed of GDM are
given by
\begin{eqnarray}
\rho_{\rm gdm}&=& \frac{1}{8M^4}(3X^2-2XM^4-M^8)~,\\
P_{\rm gdm}&=& \frac{1}{8M^4}(X-M^4)^2~,\\
c_{s}^2&=&\frac{X-M^4}{3X-M^4}~,
\end{eqnarray}
respectively. The initial condition for the value of $X(=X_0)$ is fixed at
present demanding that GDM component should be accounting for the dark
matter component observed today:
\begin{equation}
\Omega_{\rm gdm}=\frac{1}{24 M^2_{\rm Pl}H_0^2 M^4}(3X_0^2-2X_0M^4-M^8)=0.3
\end{equation}
Then the evolution of $X$ can be solved backward in time using
Eq.(\ref{eq:4.6}).
}

As we have discussed in the previous section,
we have to restrict our consideration
to the regime $\left| X- M^4 \right| / M^4 \ll 1$
and in this regime the sound speed
evolves as $c_s \propto \sqrt{|X-M^4|} \propto a^{-3/2}$.
Hence, if we look back the background evolution in $\Lambda$GDM
universe, the condition $|X-M^4|/M^4\ll 1$ might be violated at some
point in time. Let us define a critical scale factor when
the condition $\left| X- M^4 \right| / M^4 \ll 1$
is violated as
\begin{eqnarray}
\left| X(a_{\rm cr}) - M^4 \right| / M^4 =  2c_s^2(a_{\rm cr})
= 0.1~.
\end{eqnarray}
From this equation, the critical scale factor
$a_{\rm cr}$ can be obtained in terms
of $M$ as
\begin{eqnarray}
a_{\rm cr}/a_0 \simeq 1.0 \times
10^{-4}\left( {\Omega_{\rm gdm}h^2 \over 0.11} \right)^{1/3}
\left( {1.0 {\rm eV} \over M} \right)^{4/3}~.
\label{eqn:critscale}
\end{eqnarray}
Thus, it is found that for $M \gtrsim 1$ eV
the critical scale factor $a_{\rm cr}$
is smaller than the scale factor at the matter-radiation equality, $a_{\rm eq}$.
In the previous section, we have obtained the lower bound for $M$
from the background evolution as $M > 1$ eV.
Hence, if this bound is satisfied,
we can use the low energy effective theory, at least, after the matter-radiation equality.
Hereinafter, we consider the case for $M > 1$ eV.

\begin{figure}[htbp]
 \begin{center}
  \includegraphics{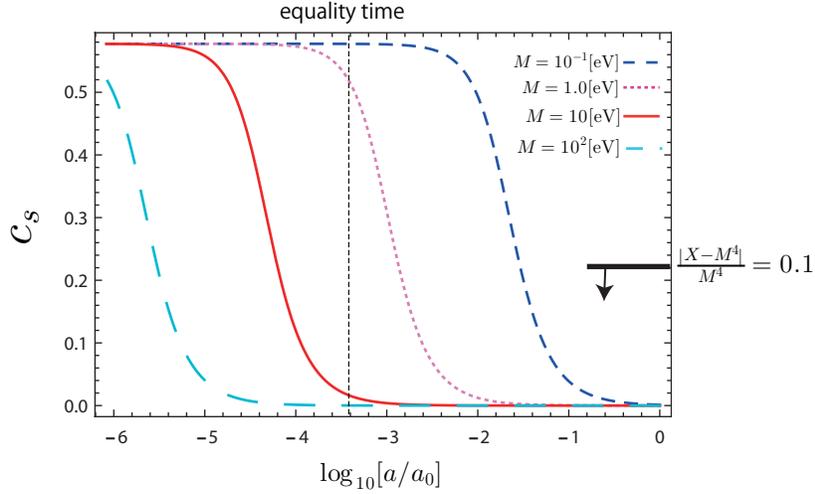}
 \end{center}
 \caption{
  The evolution of the sound speed of GDM for various values of
  ${M}$. The blue short-dashed line is for
  cosmological constant plus ghost dark matter ($\Lambda$GDM) with
  ${M}=10^{-1}$ eV, the magenta dotted line is for the $\Lambda$GDM with
  ${M}=1.0$ eV, the red solid line is for the $\Lambda$GDM with
  ${M}=10 $ eV and the cyan long-dashed line is for the $\Lambda$GDM with
  ${M}=10^2$ eV. In all plots we set $\Omega_{\Lambda}=0.7$ and
  $\Omega_{\rm gdm}=0.3$ (or $\Omega_{\rm cdm}=0.3$ for the
  $\Lambda$CDM). The vertical black dashed line represents the matter-radiation equality.
  The downward arrow shows the regime of the validity of the low energy effective theory ($\left| X - M^4\right|/M^4 < 0.1$).
  }
 \label{fig:soundspeed}
 \end{figure}

This GDM description by $X$ is equivalent to fluid-like description by
Eqs.~(\ref{eqn:rhogdm})-(\ref{eqn:GDE-GDM2}), when $X$ is near the
condensate position, i.e., $|X-M^4|\ll M^4$. However, the sound speed
will approach to $c^2_{s}=\frac{1}{3}$ in this description and will
never exceed the speed of light even if the energy density evolves to
infinity as $\left| X-M^4 \right| /M^4 \gg 1$, as is shown in
Fig.~\ref{fig:soundspeed}.  The equation of state parameter, $w=P_{\rm
gdm}/\rho_{\rm gdm}$, also approaches to $1/3$.  Therefore we can solve
the system in a numerically stable way.  It should be noted, however,
that this description should be considered only as a regulator of the
fluid description considered in the previous section, and it will be
invalid when $X$ is far away from the condensate position. Still one
needs to interpret the results where $a<a_{\rm cr}$ with caution.

In fact, we have to extend our analysis before $a_{\rm cr}$
because the density perturbations should be solved numerically from the
early epoch when the corresponding Fourier modes are outside the cosmic
horizon. Therefore, in the analysis in this section we extrapolate the
behavior of GDM beyond $a_{\rm cr}$ with a reasonable asuumption that
the sound speed of GDM had been satulated as $c_s^2=1/3$ for $a<a_{\rm
cr}$. This treatment corresponds to an assumption that the density
perturbation of GDM can not grow in the radiation dominated era even
when the low energy effective theory is broken.

 \begin{figure}[htbp]
 \begin{center}
  \includegraphics{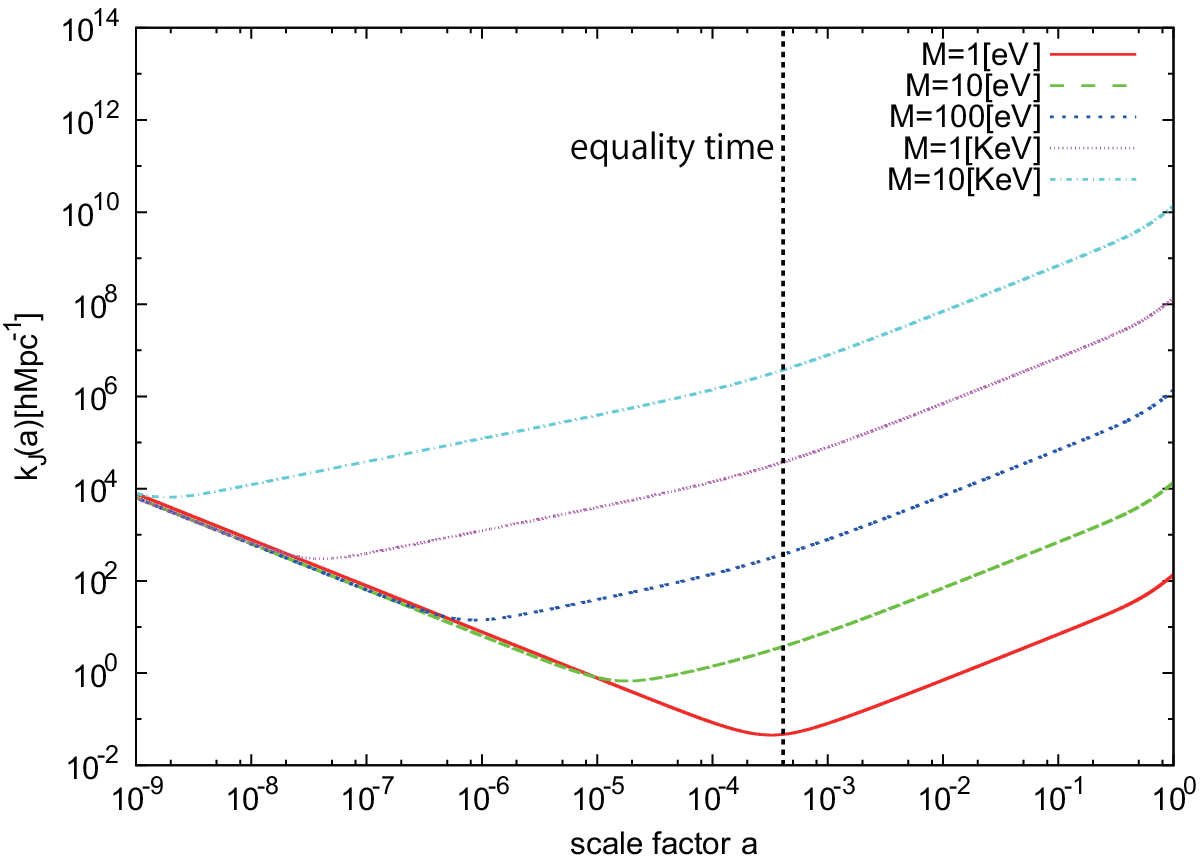}
 \end{center}
 \caption{ The comoving Jeans wavenumber for various values of ${M}$ as
  indicated in the figure. From up to bottom, the lines are for
  ${M}=10, 1 {\rm keV}, 100, 10, 1~ {\rm eV}$.  The vertical dashed line
  shows the matter-radiation equality time.  The turnover of the Jeans
  wavenumber occurs due to the change of the sound speed of GDM, which
  roughly shows the boundary of the regime of validity of the low
  energy effective theory.  }
  \label{fig:jeans}
 \end{figure}

The corresponding Jeans wavenumber is plotted in Fig.~\ref{fig:jeans}. From
the figure, we can see a turnover of the Jeans wavenumber
$k_{J}$ for each $M$.  This turnover of the
Jeans wavenumber occurs at the transition of the sound speed of the GDM
between $c_{s} =1/\sqrt{3}$ and $c_{s} \propto
a^{-3/2}$.
The Jeans wavenumber is given by
\begin{equation}
k_{J,{\rm turn}}=\frac{3}{\sqrt{2}}a_{\rm turn}H(a_{\rm turn})~,
\end{equation}
where we denote the scale factor at the turnover epoch by
$a_{\rm turn}$. We can easily find that
$a_{\rm turn} \simeq a_{\rm cr}$, where $a_{\rm cr}$ is given by
Eq.~(\ref{eqn:critscale}).
For $M \gtrsim 1~ {\rm eV}$, of course, $a_{\rm turn} \simeq a_{\rm cr}$
is smaller than $a_{\rm eq}$.

\subsubsection{Perturbation evolution}

Let us consider the evolution of the matter perturbations.
In synchronous gauge, the growth of density contrast of GDM,
$\delta_{\rm gdm}$, is described by the differential equations \cite{ma}
\begin{eqnarray}
\delta _{\rm gdm} ^{\prime \prime} &\! + \!&
\left( 1+3c_{s} ^2 -6w_{\rm gdm}\right) {\cal H}\delta_{\rm gdm} ^{\prime} \nonumber \\
&\! + \!& \Biggl\{ \frac{3}{2}{\cal H}^2( c_{s} ^2 -w_{\rm gdm})
 (1-6w_{\rm gdm}-3\bar{w}) \nonumber \\ &\!\!& \quad + c_{s} ^2 k^2
+3{\cal H}({c_{s}^2} ^{\prime}-w_{\rm gdm} ^{\prime})\Biggr\} \delta_{\rm gdm} \nonumber \\
&\! = \!& -(1+w_{\rm gdm})\frac{h^{\prime \prime}}{2}-(1+w_{\rm gdm})(1-3w_{\rm gdm}){\cal H}\frac{h^{\prime}}{2}
-\frac{w_{\rm gdm}^{\prime}h^{\prime}}{2},
\label{eqn:GDMdeq}
\end{eqnarray}
\begin{equation}
h^{\prime \prime} +{\cal H}h^{\prime} =-3{\cal H}^2(1+\bar{c_{s}}^2)\bar{\delta},
\label{eqn:einstein}
\end{equation}
where a prime denotes a derivative with respect to conformal time,
${\cal H}=a^{\prime}/a$, $h$ is the metric perturbation, $\bar{\delta}$,
$\bar{c_{s}}$, $\bar{w}$ are total component's density fluctuation,
sound speed and equation of state, respectively, defined as
\begin{eqnarray}
\bar{\delta}=\frac{\sum \rho_{i} \delta_{i}}{\sum
\rho_{i}},~~\bar{w}=\frac{\sum \rho_{i}}{\sum p_{i}}, ~~
\bar{c_s^2}=\frac{\sum \dot{\rho_i}c^2_{s, i}}{\sum \dot{\rho_i}}~.
\end{eqnarray}
The evolution of perturbations in the sound horizon depends on the time
dependence of the sound speed.  In the present model the sound speed is
$c_{s}^2=1/3$ in the very early universe, and $c_{s}^2\propto a^{-3/2}$ after some critical epoch.
Therefore, the
perturbation evolution in the sound horizon is divided into two cases.
They are; case (a): $w_{\rm gdm}=c_{s} ^2=1/3$, and case (b):
$w_{\rm gdm},c_{s} \ll 1$.

{Note that the case (a) may not be a correct description of GDM scenario
in the sense that the low
energy effective theory is broken at large redshifts where $a<a_{\rm
cr}(\lesssim a_{\rm eq})$. However, it would be natural to expect that
the density perturbation in GDM can not grow very much if the sound
speed of GDM had been saturated as $c_s\sim O (1)$.  This situation would
be effectively taken into account by setting $c_{s}^2 = \frac{1}{3}$
because in this case the density perturbation can not grow at subhorizon
scales. Therefore we shall assume case (a) in the following analysis to
solve the perturbations in GDM for the energy scales in which the
effective theory can not be applied.}

\subsubsection*{$\bullet$~case (a): $w_{\rm gdm}=c_{s} ^2=1/3$}
In this case, $c_{s} ^{\prime}, w_{\rm gdm} ^{\prime}=0$.  So,
differential equations(Eq.~(\ref{eqn:GDMdeq}), Eq.~(\ref{eqn:einstein})) are
reduced to
\begin{equation}
\delta_{\rm gdm} ^{\prime \prime} + c_{s} ^2 k^2 \delta_{\rm gdm} = -\frac{2}{3}h^{\prime \prime},
\end{equation}
\begin{equation}
h^{\prime \prime}+{\cal H}h^{\prime}=-4{\cal H}^2 \bar{\delta}.
\end{equation}
Roughly speaking, $h^{\prime \prime} \propto {\cal H}^2 \bar\delta$ from the second equation,
and using the fact that ${\cal H} \ll c_{s} k$ in sound
horizon, the first equation becomes
\begin{equation}
\delta_{\rm gdm}^{\prime \prime} + c_{s} ^2 k^2 \delta_{\rm gdm} = 0.
\end{equation}
Therefore, the density perturbation of GDM in this case cannot grow
and only oscillate.

\subsubsection*{$\bullet$~case (b) : $w_{\rm gdm},c_{s} \ll 1$}
In this case, $c_{s} ^{\prime}=-\frac{3}{2}{\cal H} c_{s}$ and $w_{\rm gdm} ^{\prime}=-3{\cal H}w_{\rm gdm}$.
So, Eq.~(\ref{eqn:GDMdeq}) becomes
\begin{equation}
\delta _{\rm gdm} ^{\prime \prime} +{\cal H}\delta _{\rm gdm} ^{\prime}
+\left( c_{s} ^2k^2 -\frac{3}{2}{\cal H}^2 \right) \delta_{\rm gdm}=0.
\end{equation}
Since ${\cal H}^{\prime}(\eta) \ll c_{s} ^{\prime} (\eta )k$, we use WKB approximation
\begin{equation}
\delta _{\rm gdm} = A(\eta ) \exp{\left( i\int ^{\eta} _{\eta _0} c_{s} kd\eta \right)}~,
\end{equation}
to solve this equation. In this approximation, $A^\prime(\eta)$ is roughly in
proportion to ${\cal H}$ and $c_{s} k \gg {\cal H}$. So the
term proportional to ${\cal H}^2$ can be neglected.  Then, the equation
becomes
\begin{equation}
\left( 2c_{s} A^{\prime} +c_{s} ^{\prime} A+c_{s} {\cal H} A\right)
ik\exp{\left( i\int ^{\eta} _{\eta _0} c_{s} kd\eta \right)}
=0.
\label{eqn:csdelta}
\end{equation}
Then we obtain $A\propto a^{1/4}$ since $c_{s} ^{\prime}=-\frac{3}{2}{\cal H}c_{s}$. Thus the growth rate of GDM for case (b) is
\begin{equation}
\delta \propto a^{1/4}~,
\end{equation}
in the sound horizon. Therefore, the density perturbation of GDM for
case (b) gradually {\it grows} in the sound horizon.
The above consideration can be generalized in the case where
$c_{s,} \propto a^{-n} (n\neq 0)$.  In this case, $c_{s}
^{\prime}=-n{\cal
H}c_{s}$, so the evolution of perturbation in sound horizon is
\begin{equation}
\delta \propto a^{(n-1)/2},
\end{equation}
which is derived from Eq.~(\ref{eqn:csdelta}).

The evolution of GDM density perturbations in $\Lambda$GDM cosmology can
be divided into four cases by comparing the wavelength of the modes and
the Jeans length.  Let us define $a_{\rm J}(k)$ as a scale factor
when a mode becomes larger than the Jeans scale, namely, when the mode
exits the sound horizon.  We define the four cases from the smallest
scale to the largest one, depending on $a_{\rm J}(k)$ and the
matter-radiation equality time denoted by $a_{\rm eq}$ as follows;

\subsubsection*{$\bullet$ case I ($k<k_{J,{\rm turn}}$):} the modes of
interest never cross the sound horizon during the cosmic history,
\subsubsection*{$\bullet$ case II ($k_{J,{\rm eq}}>k>k_{J,{\rm turn}}$):} the modes of interest exit the sound horizon
before the matter-radiation equality,
\subsubsection*{$\bullet$ case III ($a_0 > a_{\rm J}(k)$; $k>k_{J,{\rm eq}}$):}  the modes of interest exit the sound horizon after the
matter-radiation equality,
\subsubsection*{$\bullet$ case IV ($a_{\rm J}(k)>a_0$; $k>k_{J,{\rm
eq}}$):} the modes
do not exit the sound horizon until today.
\\

\begin{figure}[htbp]
\begin{center}
 \includegraphics{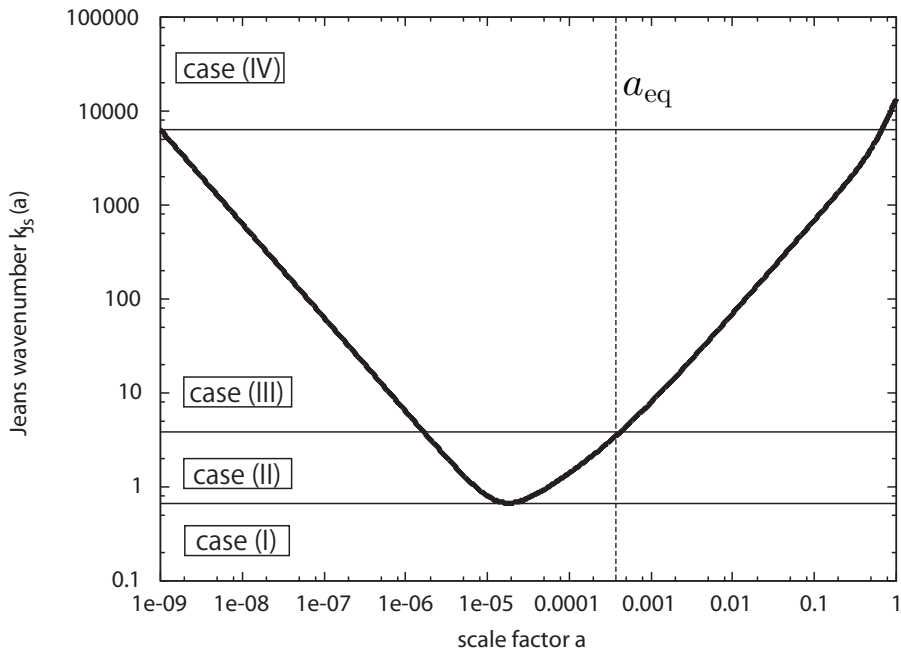}
\end{center}
\caption{Four cases (case I - IV) divided for the Jeans wavenumber for
the $\Lambda$GDM with ${M}=10 {\rm eV}$ due to the difference of the
evolution of the perturbation.  Thick solid line shows the evolution of
the Jeans wavenumber, vertical thin dashed line shows the
matter-radiation equality.}  \label{fig:fourcase}
 \end{figure}

For example, Fig.\ref{fig:fourcase} shows typical wavenumbers of four cases for the $\Lambda$GDM with
${M}=10$ eV.
If the mode belongs to the case I, the
evolution of the mode is similar to that in the ordinary CDM one because
the mode is always outside the sound horizon.  However, if the modes
belong to the cases II - IV, the modes enter the sound
horizon at some epochs, where the perturbation evolution of GDM becomes
different from that of CDM.
Then, let us consider how the modes entering the sound horizon of GDM
can be suppressed and estimate the $k$-dependence of the matter power
spectrum. In the $M \geq 1{\rm eV}$ models, the small scale
perturbations cross the sound horizon in the radiation dominated era as
shown in Fig.\ref{fig:jeans}. However, the time of exiting the sound
horizon dependends on the wavenumber of the mode $k$ and also the cut-off scale $M$ as mentioned above.

From these considerations, one can estimate the power spectrum of GDM
model. In the following, for simplicity, we neglect the phase of
oscillations in the sound horizon.  In fact, it is found to be
important quantitatively and we will include this effect in the
numerical calculation which will be presented later in order to
calculate evolution of perturbations and the matter power spectrum more
precisely.  We approximate $\delta_{\rm gdm} \propto a$ and $\propto
\log a$ outside the sound horizon in the matter and radiation dominated
era ($a_{\rm cr}<a<a_{\rm eq}$), respectively. Because of the fact that $\delta_{\rm gdm} \propto
a^{1/4}$ inside the sound horizon when $a \gtrsim a_{\rm cr}$, we find
\begin{eqnarray}
\delta _{\rm gdm}(a_{0})\sim \left\{ \begin{array}{ll}
\delta _{\rm cdm}(a_{0}) &  \mbox{for}~~ k<k_{J,{\rm turn}} \\
\left( {a_{\rm J}}/{a_{\rm cr}}\right) ^{1/4} \log\left({a_{\rm
 eq}}/{a_J}\right)\left({a_0}/{a_{\rm eq}}\right)\delta _{\rm
 gdm}(a_{\rm hc}) & \mbox{for}~~ k_{J,{\rm eq}}>k>k_{J,{\rm turn}} \\
\left( {a_{\rm J}}/{a_{\rm cr}}\right) ^{1/4} \left({a_0}/{a_J}\right)\delta _{\rm gdm}(a_{\rm hc}) & \mbox{for}~~
{a_0 > a_{\rm J}(k);~ k>k_{J,{\rm eq}}} \\
\left( {a_{0}}/{a_{\rm cr}}\right) ^{1/4} \delta
 _{\rm gdm}(a_{\rm hc}) & \mbox{for}~~
{a_{\rm J}(k)>a_0;~ k>k_{J,{\rm eq}}} \\
\end{array} \right. ,
\end{eqnarray}
where we have introduced $a_{\rm hc}$ as the scale factor when the modes of interest enter in the Hubble horizon.
In these equations, only $a_{\rm J}$ has $k$ dependence, namely, $a_{\rm J}
\propto k^2$ in the radiation era and $a_{\rm J} \propto k$ in the matter era.
Meanwhile, in the standard cold dark matter cosmology, we find $\delta
_{\rm cdm}(a_{0})\sim \log({a_{\rm eq}}/{a_{hc}})\left({a_0}/{a_{\rm eq}}\right) \delta _{\rm gdm}(a_{\rm hc})$.
Therefore, the $k$ dependence of the GDM power spectrum at present can be found
to be
\begin{eqnarray}
\frac{P_{\rm gdm}(k)}{P_{\rm cdm}(k)} \propto \left\{ \begin{array}{ll}
1 &  \mbox{for}~~ k<k_{J,{\rm turn}}\\
\left(\log k \right)^{-2} & \mbox{for}~~
k_{J,{\rm eq}}> k>k_{J,{\rm turn}}\\
k^{-2} &  \mbox{for}~~
{a_0 > a_{\rm J}(k);~ k>k_{J,{\rm eq}}}\\
(\log k)^{-2} & \mbox{for}~~
{a_{\rm J}(k)>a_0;~ k>k_{J,{\rm eq}}}\\
\end{array} \right. .
\end{eqnarray}
where we have used $x^{-1/4}\log(x)\approx O(1)$ for $x>1$.

The analysis so far is based on the rough estimate and only
appropriate for qualititative understanding. In order to evaluate the
perturbation amplitude quantitatively and take into account the effects
neglected in the above analysis, we calculate the evolution numerically
using the modified CAMB code \cite{camb}.

\begin{figure}[htbp]
 \begin{center}
   \includegraphics{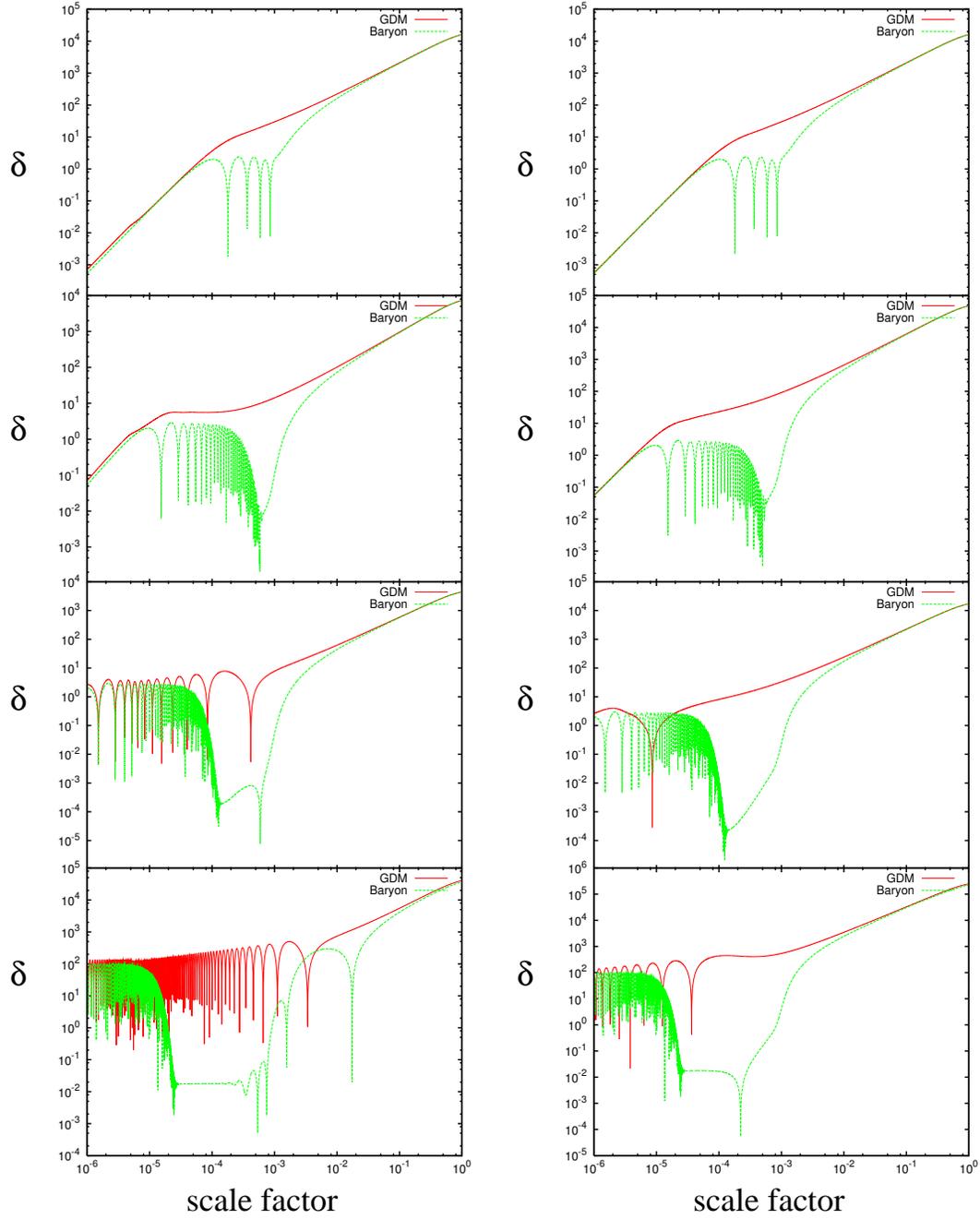}
 \end{center}
 \caption{The evolution of density perturbations of GDM and baryon fluid
 in Ghost condensation model.  Left panels are for $M=20~{\rm eV}$ model,
 right panels are for $M=100~{\rm eV}$ model.  The panels from top to
 bottom are for $k=0.1, 1, 10, 100~{\rm Mpc^{-1}}$.} %
\label{fig:variable plot}
\end{figure}

In Fig.\ref{fig:variable plot}, we depict the evolutions of density
perturbations of GDM with different wavenumbers ($k=0.1, 1, 10, 100$
Mpc$^{-1}$) for two different model parameters ($M=20$~eV and
$100$~eV). We also depict those of the baryon density for comparison. At
largest scales (top panels) the evolution of GDM density perturbations
is almost identical to that of standard CDM, because the mode is always
outside the sound horizon. On the other hand, the evolutions of the
modes at small scales are depicted in the third and fourth panels. In
these panels the amplitude of GDM exhibits oscillations when the
perturbations are inside the
sound horizon. The growth of density perturbation of GDM inside the sound
horizon for $w_{\rm gdm}, c^2_{s} \ll 1$ is clearly seen, which has a
dependence of $a^{1/4}$ as derived in our analytic estimate.  We need,
however, to see the second-top left panel ($k=1$~Mpc$^{-1}$ and $M=20$
eV) with care. The panel shows the marginal case where the mode exits
the sound horizon immediately after entering the horizon. In this case,
we observe that the amplitude of $\delta_{\rm gdm}$ does not experience
any logarithmic growth which is expected for the CDM case. We found that
this brings a large difference in amplitudes at the present universe about
an order of magnitude, which is seen by comparing with the second-top right
panel.

\begin{figure}[ht]
\begin{center}
\includegraphics[keepaspectratio=true,height=70mm]{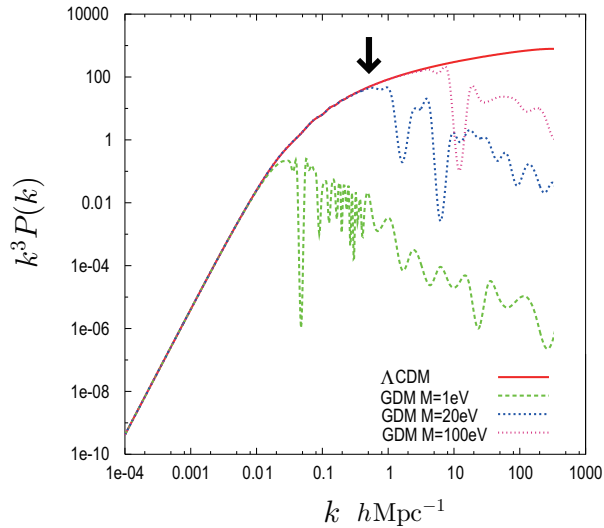}
\end{center}
\caption{The matter power spectrum for GDM model.  Red solid line is a
matter power spectrum of $\Lambda$CDM model.  The GDM power spectra are
plotted for three models : $M=1~{\rm eV}$ (green dashed line), $M=20~{\rm
eV}$ (blue short-dashed line), $M=100~{\rm eV}$ (pink dotted line).  For modes
$k>k_{J,{\rm turn}}$ (shown by a down arrow for $M = 20~{\rm eV}$), the power spectrum of GDM model is suppressed compared to that of
$\Lambda$CDM model.}  \label{fig:GDMpower}
\end{figure}

The net effect in the matter power spectrum is the {\it deficit} of
power for $k \gtrsim k_{J,{\rm turn}}$, which is shown in
Fig.\ref{fig:GDMpower}.
From this figure, we can find that
the matter power spectrum alters
not at $k_{J,{\rm eq}}$ as discussed in Sec.4.1, but at $k_{J,{\rm
turn}}$.
That is because we have extended the analysis beyond the critical epoch
$a_{\rm cr}$ in our numerical calculations, by extrapolating the
behavior of GDM beyond $a_{\rm cr}$ with a reasonable assumption that
the sound speed of GDM had been satulated as $c_s^2=1/3$.
In this case
the density perturbation of GDM only oscillates inside the sound horizon
and does not experience any logarithmic growth during
the radiation dominated era
which is expected in the
$\Lambda$CDM model.  The power deficit at $k\gtrsim k_{J,{\rm eq}}$ is
more reasonable because it is a result derived mainly from the
perturbation evolution at the subhorizon scales at the regime where the
effective low-energy theory is applicable, i.e., $(X-M^4)/M^4 < 1$ or
$a>a_{\rm cr}$, under a simple assumption that the density perturbation
of GDM can not grow deep in the radiation dominated era, even when the
low energy effective theory is broken.

The evolution of the baryon density perturbation is affected only
through gravity. In the standard cosmology, it is known that after the
fluctuation is erased by diffusion damping the baryon fluctuation should
start to grow in time again with terminal velocity falling into the
gravitational potential of clustered dark matter
\cite{Yamamoto:1997qc}. In the GDM model, however, the baryon fluid can not
have terminal velocity because the dark matter density fluctuation
oscillates in the sound horizon and hence the gravitational potential.
As the GDM ceases oscillating, the amplitude of the density fluctuation of
baryon fluid starts growing with oscillation, which is clearly seen in the
left bottom panel. In any case, the baryon density fluctuation quickly
catches up the dark matter density one after decoupling.

The matter power spectrum of our universe can be probed through various
cosmological observations, such as clustering of galaxies
\cite{Tegmark:2006az}, cosmic shear
\cite{Fu:2007qq,Benjamin:2007ys,Schrabback:2009ba}, Lyman-$\alpha$
forest \cite{Croft:2000hs,Kim:2003qt,McDonald:2004xn}, and so on. At
present, almost all observational data support the CDM paradigm, roughly
for the wavenumber $k/h \lesssim 1$ Mpc$^{-1}$. In the models considered in
this paper, this observational fact gives us a constraint on the model
parameter $M$. The constraint is derived through the suppression of the
matter power spectrum, in the same way to obtain the constraints on the
hot and/or warm dark matter models (or in other words, masses of
neutrinos and/or warm dark matter particles) \cite{Boyarsky:2008xj}. By
looking at the matter power spectrum obtained in the present analysis
(Fig.\ref{fig:GDMpower}), we conclude that the model parameter should be
\begin{equation}
M\gtrsim 10 ~\mbox{eV}~,
\end{equation}
which is a stronger constraint
than that obtained only from the background evolution considered in
Sec. \ref{sec:background}.
This result is consistent with the rough analytic estimate given by Eq.~(\ref{eqn:critwavenumber}).

Finally we should note that the way of suppression in the GDM models
is different from that in hot or warm dark matter models. In the GDM
models the suppression is the power law while for hot and warm models
the power is exponentially suppressed. This fact might be used to
distinguish between GDM and other dark matter models.

\section{Summary and discussion}

We have investigated the possibility that ghost condensation may serve
as an alternative to dark matter, which had not been investigated in detail.
In the present paper we have considered the $\Lambda$GDM
universe, i.e. a late-time universe dominated by a cosmological constant
and ghost dark matter. We have investigated the
Friedmann-Robertson-Walker (FRW) background evolution and the
large-scale structure of the $\Lambda$GDM universe, and have found a
lower bound on the scale of spontaneous Lorentz breaking as
$M\gtrsim 10~{\rm eV}$. This bound is compatible with the
phenomenological upper bound $M\lesssim 100~{\rm GeV}$ known in the
literature.

As we have reviewed in Sec.~\ref{sec:review}, ghost condensation is the
simplest Higgs mechanism for gravity in the sense that the number of
Nambu-Goldstone boson is only one. The structure of the low-energy
effective field theory is completely determined by the symmetry breaking
pattern and this makes it possible for us to give robust predictions of
the theory as far as the system is in the regime of validity of the
effective theory. In Sec.~\ref{sec:review} we have also reviewed the
infrared modification of linearized gravity as well as some non-linear
dynamics and the phenomenological lower bound mentioned above.

In Sec.~\ref{sec:background} we have provided a simple description of
ghost dark matter and investigated the FRW background evolution.
In subsection~\ref{subsec:simple-gdm} we have shown that, under a
certain condition, the background evolution and the behavior of
large-scale perturbations of ghost dark matter can be described by a
fluid with the equation of state
$P_{\rm gdm}\propto\rho_{\rm gdm}^2/M^4$.
This description has been used throughout this paper, except for
numerical studies where the field picture turns out to be more
convenient for some technical reasons. In
subsection~\ref{subsec:gi2gdm}, as a possible cosmological scenario
relating the late time evolution of ghost dark matter to the early
universe, we have considered a generation mechanism of ghost dark matter
at the end of ghost inflation~\cite{ArkaniHamed:2003uz}. In
subsection~\ref{subsec:background} we have shown that the background
FRW evolution in the $\Lambda$GDM universe is indistinguishable from
that in the standard $\Lambda$CDM universe if $M\gtrsim 1~{\rm eV}$.

In Sec.\ref{sec:LSS} we have investigated the large-scale structure of
the $\Lambda$GDM universe. Since the GDM has the effective sound speed
unlike the standard CDM, small scale perturbations are suppressed. The
suppression of the matter power spectrum occurs in a way different from
that in models with hot dark matter or warm dark matter particles. We
have given an analytic treatment to predict the matter power spectrum
observed today and have also calculated the power spectrum
numerically.  The analytic treatment can be extended for general dark
matter models in which dark matter has a finite sound speed whose time
dependence is the power law of scale factor.  By comparing the GDM power
spectrum with the observed one, one can get a lower bound on the scale
$M$. The result we obtained is $M\gtrsim 10~{\rm eV}$.

The constraint obtained in this paper can be improved further by
observations of the matter power spectrum at smaller scales. For example,
21cm-line observations and/or the precise determination of the
reionization epoch will provide us plenty of information about the
matter power spectrum at smaller scales and hence stronger limits on
the ghost condensation scale $M$. We leave these interesting subjects for
future investigation.

As suggested in \cite{ArkaniHamed:2003uy}, ghost condensation may
provide an alternative explanation for the acceleration of the present
universe if $M\sim 10^{-3}~{\rm eV}$. If the cosmological constant in the
symmetric phase (with $\dot{\phi}=0$) is zero then the effective
cosmological constant in the broken phase, i.e. the ghost condensate,
(with $\dot{\phi}=M^2$) is positive~\footnote{Note that in the
standard Higgs mechanism, the effective cosmological constant in the
broken phase would be negative if the cosmological constant in the
symmetric phase is zero.} and of order $O(M^4/M_{\rm Pl}^2)$ unless
fine-tuned. Unfortunately, the condition $M\sim 10^{-3}~{\rm eV}$ is not
compatible with the lower bound on $M$ found in the present paper under
the assumption that ghost dark matter is responsible for all dark matter
in the universe. Therefore, it is not easy for the ghost condensate to be a
simultaneous alternative to dark energy and dark matter unless
fine-tuned.

On the other hand, as shown in subsection~\ref{subsec:gi2gdm}, ghost
inflation is compatible with ghost dark matter. Ghost dark matter is
naturally produced at the end of ghost inflation. Moreover, the lower
bound (\ref{eqn:bound-inflation}) from ghost inflation can be satisfied
simultaneously with not only the phenomenological upper bound
(\ref{eqn:upperbound}) but also the bound from ghost dark
matter. Detailed investigation of the combination of ghost inflation and
ghost dark matter is certainly worthwhile as a future work.

\begin{acknowledgments}
 We thank T. Chiba for informing us of a related work
 Ref.~\cite{Giannakis:2005kr}.
 This research was supported by the Grant-in-Aid for Nagoya University
 Global COE Program, "Quest for Fundamental Principles in the Universe:
 from Particles to the Solar System and the Cosmos", by Grant-in-Aid for
 Scientific Research 21740177 (KI),
 17740134 (SM), 19GS0219 (SM), 21111006 (SM), 21540278 (SM),
 by World Premier International Research Center Initiative (WPI
 Initiative),
 and by Grant-in-Aid for Scientific Research on Priority Areas No. 467
 ``Probing the Dark Energy through an Extremely Wide and Deep Survey
 with Subaru Telescope'', from MEXT of Japan.
\end{acknowledgments}

\end{document}